\documentclass{PoS}
\usepackage{bm}
\usepackage{multirow}
\usepackage{chemarrow}
\usepackage{amsmath}
\usepackage{graphicx}
\usepackage{subfigure}

\title{Two-Baryon Systems with Twisted Boundary Conditions}

\ShortTitle{Two-Baryon Systems with Twisted Boundary Conditions}

\author{\speaker{Zohreh Davoudi}%
         \thanks{In collaboration with Ra\'ul A. Brice\~no, Thomas C. Luu and Martin J. Savage.}\\
        Center for Theoretical Physics, Massachusetts Institute of Technology, Cambridge, MA 02139, USA\\
        E-mail: \email{davoudi@mit.edu}}

\abstract{I derive the most general quantization condition for energy eigenvalues of two interacting baryons in a finite cubic volume when arbitrary twisted boundary conditions are imposed on their finite-volume wavefunctions. These quantization conditions are used, along with experimentally known scattering parameters of two-nucleon systems in the coupled $^{3}S_1-{^{3}}D_1$ channels, to demonstrate the expected effect of a selection of twisted boundary conditions on the spectrum of the deuteron. It is shown that an order of magnitude reduction in the finite-volume corrections to the deuteron binding energy arise in moderate volumes with a proper choice of boundary conditions on the proton and the neutron, or by averaging the result of periodic and anti-periodic boundary conditions. These observations mean that a sub-percent accuracy can be achieved in the determination of the deuteron binding energy at (spatial) volumes as small as $\sim(9~[\rm{fm}])^3$ in upcoming lattice QCD calculations of this nucleus with physical light-quark masses. The results reviewed in this talk are presented in details in Ref. \cite{Briceno:2013hya}.}

\FullConference{The 32nd International Symposium on Lattice Field Theory\\
                 23-28 June, 2014\\
                 Columbia University New York, NY}

\begin{document}

\section{Introduction}
With the aid of algorithmic and computational advances, lattice quantum chromodynamics (LQCD) studies of properties of light nuclei and hyper nuclei have become reality in recent years (albeit still with unphysical quark masses) \cite{Beane:2012vq, Yamazaki:2012hi, Beane:2014ora}. These are starting to pave the road towards truly \emph{ab initio} calculations in nuclear physics which ultimately aim to reduce the systematic uncertainties associated with several crucial nuclear observables \cite{Briceno:2014tqa}. LQCD calculations themselves suffer from systematic uncertainties. These are however fully quantifiable, and can be reduced by an increase in computational resources and/or with the help of ongoing formal developments \cite{Beane:2014oea}. Among these systematics are those arising from the finite  extents of space-time in these calculations.\footnote{We take the temporal extent of the volume to be infinite in this study. In reality the finite-time (i.e. thermal) effects due to backward propagating states must be taken into account, and be quantified, in any LQCD calculation.} The finite-volume (FV) corrections to the mass of single hadrons are fully governed by the finite-range of hadronic interactions, and are known to be exponential in the spatial extent of the volume, $\rm L$, and the mass of the pion, $m_{\pi}$, scaling as $e^{-m_{\pi} {\rm L}}/{\rm L}$ at leading order (LO) \cite{Luscher:1985dn}. For the two-hadron bound states,  due to the introduction of a new scale in the system, i.e., the binding momentum of the bound states, $\kappa$, further volume dependences arise in the calculated energies, which at LO scale as $e^{-\kappa {\rm L}}/{\rm L}$. For near-threshold bound states such as the deuteron, these latter corrections can be rather large even in small to intermediate volumes at the physical point -- corresponding to the physical light-quark masses (see Sec. \ref{deuteron} for a quantitative demonstration of this fact in the deuteron system). This talk aims to review a recent proposal in Ref. \cite{Briceno:2013hya} that leads to a considerable reduction of such volume effects with judicious choices of boundary conditions (BCs) in the two-hadron systems in a finite cubic volume. Other (closely related) strategies such as introducing multiple center-of-mass (CM) boosts to the two-nucleon systems can be found in, e.g., Refs. \cite{Davoudi:2011md, Briceno:2013bda}.

The most commonly used lattice geometries are (hyper) cubic and the boundary conditions that are generally imposed on gauge and quark fields are periodic (anti-periodic for quark fields in the temporal direction). These BCs can be simply generalized to twisted BCs (TBCs) by allowing the quark fields to acquire a non-unity phase at the boundaries, $\psi(\mathbf{x}+\mathbf{n}{\rm{L}})=e^{i{\rm{\theta}} \cdot \mathbf{n}}\psi(\mathbf{x})$, characterized by a twist angle $0<\theta_i<2\pi$ in the $i^{\rm th}$ Cartesian direction, with $\mathbf{n}$ being an integer triplet. This allows the (free) single-hadron momentum modes to take non-integer values, $\mathbf{p}=\frac{2\pi}{{\rm L}}+\frac{\bm{\phi}}{{\rm L}}$ where $\bm{\phi}$ is the corresponding twist angle of the hadron due to the TBCs imposed on its (valence) quark-level interpolator. These BCs were first introduced in the (S-wave) two-nucleon systems in Ref. \cite{Bedaque:2004kc} as a knob to shift the location of FV energy levels\footnote{The effectiveness of this method in two-hadron systems with identical hadrons, or with equal twist angles imposed on each hadron, is lost as discussed in this review. This point must be be kept in mind when consulting Ref. \cite{Bedaque:2004kc}.} and therefore provide further kinematic inputs to a modified L\"usher-type quantization condition (QC) \cite{Luscher:1986pf, Luscher:1990ux} that gives access the scattering amplitudes. It was shortly realized that such BCs can as well be used in LQCD calculations of transition matrix elements to resolve the threshold region without requiring large lattice volumes, e.g., \cite{Tiburzi:2005hg, Jiang:2006gna}. In this review I focus on one aspect of these BCs, namely their role in improving the volume dependence of binding energies in the two-hadron systems with an emphasis on the deuteron. I devote Sec. \ref{QC} of this review to sketch a derivation of the most general FV QC for two-baryon systems with arbitrary TBCs as we first presented in Ref. \cite{Briceno:2013hya}. This QC is used in Sec. \ref{deuteron} to investigate the volume effects on the deuteron binding energy at the physical point, and has enabled us to identify the twist angles, or combinations thereof, that give rise to significant cancellations of volume effects. I conclude by several comments in Sec. \ref{discussions}.

\section{Finite-volume Quantization Condition
\label{QC}}
The most general QC for two spin-$\frac{1}{2}$ baryons can be obtained from that of two nucleons as derived in Ref. \cite{Briceno:2013lba},\footnote{The case of two spin-$0$ particles as well as a spin-$0$ and a spin-$\frac{1}{2}$ particle can be obtained from the result presented in this section by inputting the corresponding values of spin in Eq. (\ref{deltaG}). For two hadrons with arbitrary spin the most general QC is presented in Ref. \cite{Briceno:2014oea}.} upon accounting for relativistic kinematics that will be considered here, as well as modifications due to the use of TBCs. These only alter the on-shell kinematics of the system, and therefore modify the allowed non-interacting momentum modes in the volume.\footnote{In deriving this QC, the exponential corrections of the form $e^{-m_{\pi}{\rm L}}$ -- that are arising from the finite range of hadronic interactions -- are neglected.} In order to identify these modes, let us consider the two-hadron wavefunction in the lab frame $\psi_{\rm Lab}$~\cite{Rummukainen:1995vs, Davoudi:2011md}. This wavefunction is subject to the TBCs,
\begin{eqnarray}
\psi_{{\rm Lab}}(\mathbf{x}_1+{\rm{L}}\mathbf{n}_1,\mathbf{x}_2+{\rm{L}}\mathbf{n}_2)
\ =\ 
e^{i{\bm{\phi}}_1 \cdot \mathbf{n}_1+i{\bm{\phi}}_2 \cdot \mathbf{n}_2}
\ \psi_{{\rm Lab}}(\mathbf{x}_1,\mathbf{x}_2),
\label{WF-BC}
\end{eqnarray}
where $\mathbf{x}_1$ and $\mathbf{x}_2$ denote the position of two hadrons,
 $\bm{\phi}_1$  and $\bm{\phi}_2$ are their respective twist angles, 
and $\mathbf{n}_1,\mathbf{n}_2\in\mathbb{Z}^3$.  Denoting the total (conserved) momentum of the system by $P=(E,\mathbf{P})$ in the lab frame, the equal-time wavefunction of the system is
\begin{eqnarray}
\psi_{{\rm Lab}}(x_1,x_2) 
\ =\  e^{-iE X^0+i\mathbf{P} \cdot \mathbf{X}}
\ \varphi_{{\rm Lab}}(0,\mathbf{x}_1-\mathbf{x}_2),
\end{eqnarray}
where $X$ is the position of the CM, $X =  \alpha x_1+(1-\alpha) x_2$ with 
$\alpha=\frac{1}{2}\left(1+\frac{m_1^2-m_2^2}{E^{*2}}\right)$ for particles of masses $m_1$ and $m_2$ \cite{Davoudi:2011md}, and where the total CM entergy of the systemm is $E^*=\sqrt{E^2-\mathbf{P}^2}$.  
Since the CM wavefunction is independent of the relative time coordinate~\cite{Rummukainen:1995vs}, one has  
$\varphi_{{\rm Lab}}(0,\mathbf{\mathbf{x}_1-\mathbf{x}_2})=\varphi_{{\rm CM}}(\hat{\gamma} (\mathbf{x}_1-\mathbf{x}_2))$,
where the boosted vectors are defined as 
$\hat{\gamma} \mathbf{x}=\gamma \mathbf{x}_{\Vert}+ \mathbf{x}_{\bot}$ with $\gamma=E/E^*$. $\mathbf{x}_{\Vert}$ ($\mathbf{x}_{\bot}$) is the component of $\mathbf{x}$ that is 
parallel (perpendicular) to $\mathbf{P}$. 
By writing Eq.~(\ref{WF-BC}) in terms of $\varphi_{CM}$ it follows that
\begin{eqnarray}
e^{i\alpha\mathbf{P}\cdot(\mathbf{n}_1-\mathbf{n}_2){\rm L}+i\mathbf{P}\cdot\mathbf{n}_2{\rm L}}
\ \varphi_{{\rm CM}}(\mathbf{y}^*+\hat{\gamma}(\mathbf{n}_1-\mathbf{n}_2){\rm L})
\ =\ 
e^{i{\bm{\phi}}_1 \cdot \mathbf{n}_1+i{\bm{\phi}}_2 \cdot \mathbf{n}_2}
\ \varphi_{{\rm CM}}(\mathbf{y}^*),
\end{eqnarray}
where $\mathbf{y}^*=\mathbf{x}_1^*-\mathbf{x}_2^*$ is the relative coordinate of two hadrons in the CM frame. Now one can take the Fourier transform of this relation to arrive at  
the system's allowed relative momenta, 
\begin{eqnarray}
\mathbf{r}
\ =\ 
\frac{1}{{\rm L}}\ 
\hat{\gamma}^{-1}
\ \left[2\pi(\mathbf{n}-\alpha\mathbf{d})-(\alpha-\frac{1}{2})(\bm{\phi}_1+\bm{\phi}_2)+\frac{1}{2}(\bm{\phi}_1-\bm{\phi}_2)\right],
\label{r-TBC}
\end{eqnarray}
where we have used $\mathbf{P}=\frac{2\pi}{{\rm L}}\mathbf{d}+\frac{\bm{\phi}_1+\bm{\phi}_2}{{\rm L}}$ 
with $\mathbf{d}\in \mathbb{Z}^3$. The sum over momentum modes in the QC is a sum over integer vectors $\mathbf{n}\in\mathbb{Z}^3$ in Eq. (\ref{r-TBC}). This QC can be written in general as a determinant condition
\begin{eqnarray}
\det\left[{(\mathcal{M}^{\infty}(p^*))^{-1}+\delta\mathcal{G}^{V}(p^*)}\right]\ =\ 0,
\label{TBC-QC}
\end{eqnarray}
where $\mathcal{M}^{\infty}$ is the infinite-volume scattering amplitude and $\delta\mathcal{G}^V$ is a FV matrix whose elements in the total angular momentum basis can be written as 
\begin{eqnarray}
&& \left[\delta\mathcal{G}^V\right]_{JM_J,LS;J'M_J',L'S'}=i	\eta \frac{p^*}{8\pi E^*}
\delta_{SS'}\left[\delta_{JJ'}\delta_{M_JM_J'}\delta_{LL'} +i\sum_{l,m}\frac{(4\pi)^{3/2}}{p^{*l+1}}
c_{lm}^{\mathbf{d},\bm{\phi}_1,\bm{\phi}_2}(p^{*2};{\rm L}) \right.
\nonumber\\
&& \qquad \qquad \qquad \left .  \times \sum_{M_L,M_L',M_S}\langle JM_J|LM_L,SM_S\rangle \langle L'M_L',SM_S|J'M_J'\rangle 
\int d\Omega~Y^*_{L M_L}Y^*_{l m}Y_{L' M_L'}\right].
\label{deltaG}
\end{eqnarray}
%
$p^*$ is the momentum of each particle in the CM frame. $\eta=1/2$ for identical particles and $\eta=1$ otherwise. $\langle JM_J|LM_L,SM_S\rangle$ are Clebsch-Gordan coefficients, where $J$ is the total angular momentum, $M_J$ is the eigenvalue of the $\hat J_z$ operator, and $L$ and $S$ are the orbital angular momentum and the total spin of the system, respectively. The volume dependence and the dependence on the BCs show up in the kinematic functions 
 $c_{lm}^{\mathbf{d},\bm{\phi}_1,\bm{\phi}_2}(p^{*2};{\rm L})$, 
 defined as
 \begin{eqnarray}
c^{\textbf{d},\bm{\phi}_1,\bm{\phi}_2}_{lm}(p^{*2};{\rm L})
\ =\ \frac{\sqrt{4\pi}}{\gamma {\rm L}^3}\left(\frac{2\pi}{{\rm L}}\right)^{l-2}
\mathcal{Z}^{\mathbf{d},\bm{\phi}_1,\bm{\phi}_2}_{lm}[1;(p^*{\rm L}/2\pi)^2],
\label{clm}
\end{eqnarray}
with
\begin{eqnarray}
\mathcal{Z}^{\mathbf{d},\bm{\phi}_1,\bm{\phi}_2}_{lm}[s;x^2]
\ =\ \sum_{\mathbf r}
\frac{ |{\bf r}|^l \ Y_{l m}(\mathbf{r})}{(\mathbf{r}^2-x^2)^s},
\label{Zlm}
\end{eqnarray}
where we already have determined the momentum modes $\mathbf{r}$ to be used in the FV sums in Eq. (\ref{r-TBC}). This completes the derivation of the QC in Eq. (\ref{TBC-QC}).
The QCs corresponding to PBCs \cite{Rummukainen:1995vs, Davoudi:2011md} can be straightforwardly obtained from Eqs. (\ref{TBC-QC}) and (\ref{deltaG}) by setting $\bm{\phi}_1=\bm{\phi}_2=\mathbf{0}$. 
Our result also recovers two limiting cases that are considered in Ref.~\cite{Agadjanov:2013kja} for the use of TBCs in the scalar sector of QCD. 
As is clear from Eq. (\ref{r-TBC}), for particles of equal mass (e.g., the np system in the isospin limit), imposing the same TBCs on the particles (e.g., corresponding to imposing the same twist on the u and d quarks in the np system) will eliminate the dependence of the CM energy on the twist angles. This means that one can not shift the location of the CM energy levels by changing the twist angle.

\section{Deuteron Spectra and Volume-effects Improvement
\label{deuteron}}
The volume-dependence of the binding energies can be obtained from the QC derived in Sec. \ref{QC} by performing an analytic continuation in momentum, $p^*=i\kappa$, where $\kappa$ denotes the binding momentum of a bound state with binding energy $B$ (in the non-relativistic limit, $B=-(E^*-m_1-m_2) \approx {\kappa^2}/{2\mu}$ where $\mu$ denotes the reduced mass of the system). To quantitatively illustrate the effect of BCs on the energies, we focus on the deuteron, the lightest nucleus at the physical point with a binding energy of $B_d^{\infty}=2.224644(34)~{\rm MeV}$. Due to the negligible scattering in higher partial waves at low energies, the infinite-dimensional QC in Eq. (\ref{QC}) can be truncated to a finite space where only the contributions from scattering channels with $L \leq 2$ are taken into account. This QC will be used in upcoming studies of two-nucleon systems with TBCs to constrain the scattering parameters in these channels, including the $S-D$ mixing parameter. However, here we take the fits to experimentally known phase shifts and mixing parameters -- when extrapolated to negative energies \cite{Briceno:2013bda} -- to in turn predict the expected spectrum in a finite volume with a selection of TBCs. In particular, we choose the following twist angles:  $\bm{\phi}^p=-\bm{\phi}^n \equiv \bm{\phi}=(0,0,0)$ (PBCs), 
$(\pi,\pi,\pi)$ (APBCs) and $(\frac{\pi}{2},\frac{\pi}{2},\frac{\pi}{2})$ (i-PBCs), which at the level of the quarks, this implies that the twist angles of the (valence) up and down quarks are ${\bm\phi}^u =  -{\bm\phi}^d = {\bm\phi}$. We also set $\mathbf{d}=\mathbf{0}$ in Eq.~(\ref{r-TBC}) so that the np system is at rest in the lab frame. 

\begin{figure}[h]
\begin{center}
\includegraphics[scale=0.345]{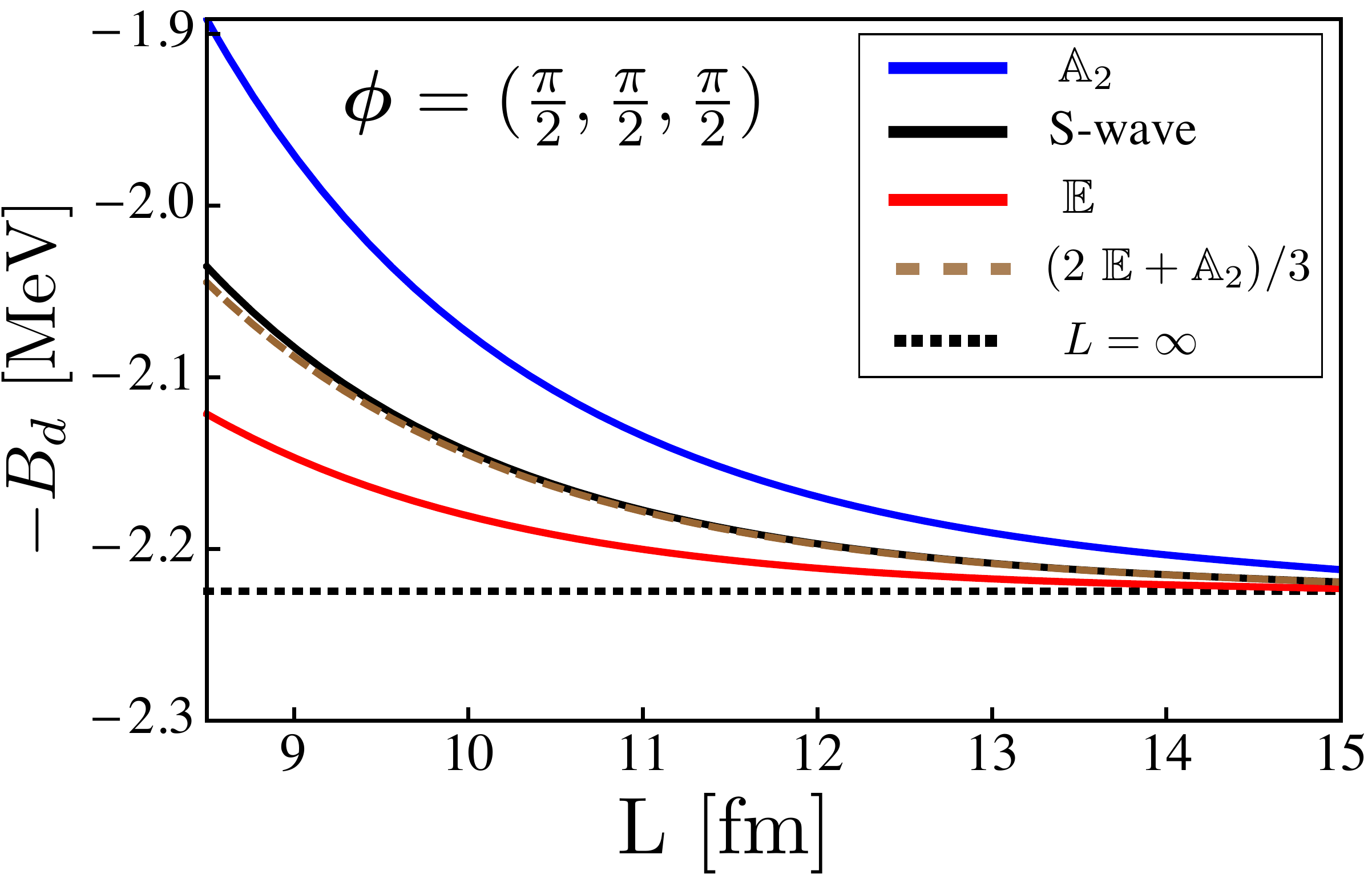}
\caption{
The deuteron binding energy as a function of ${\rm L}$ 
using i-PBCs
($\bm{\phi}^p=-\bm{\phi}^n \equiv \bm{\phi}=(\frac{\pi}{2},\frac{\pi}{2},\frac{\pi}{2})$). 
The blue curve corresponds to the $\mathbb{A}_2$ irrep of the $C_{3v}$ group, 
while the red curve corresponds to the $\mathbb{E}$ irrep. 
The brown-dashed curve corresponds to the weighted average of the $\mathbb{A}_2$ and 
$\mathbb{E}$ irreps, $-\frac{1}{3}(2B_d^{(\mathbb{E})}+B_d^{(\mathbb{A}_2)})$,
while the black-solid curve corresponds to the S-wave limit.
The infinite-volume deuteron binding energy is shown by the black-dotted line. The figure is reproduced from Ref. \cite{Briceno:2013hya}.
}
\label{fig:A2-E}
\end{center}
\end{figure}

The energy levels obtained from the QC in Eq. (\ref{TBC-QC}) correspond to different irreducible representations (irreps) of the point symmetry group of the FV calculation. For the case of i-PBCs, the symmetry group is $C_{3v}$ and the first lowest-lying energy levels correspond to the two-dimensional irrep $\mathbb{E}$ and the one-dimensional irrep $\mathbb{A}_2$ of the $C_{3v}$ group, respectively. As is seen from Fig. \ref{fig:A2-E}, at volumes as small as $\sim (9~[{\rm fm}])^3$, the FV energies in both irreps are already very close to the infinite-volume value.\footnote{This is almost the smallest volume that can be used in LQCD calculations of two-hadron systems to assure the neglected exponential corrections to the L\"uscher QC (of the form $e^{-m_{\pi}{\rm L}}$) are below percent level at the physical point.} The spin-averaged energy, defined as $-\frac{1}{3}(2B_d^{(\mathbb{E})}+B_d^{(\mathbb{A}_2)})$, provides even better agreement with the infinite-volume value with negligible volume effects.

\begin{figure}[h]
\begin{center}
\subfigure[]{
\includegraphics[scale=0.3]{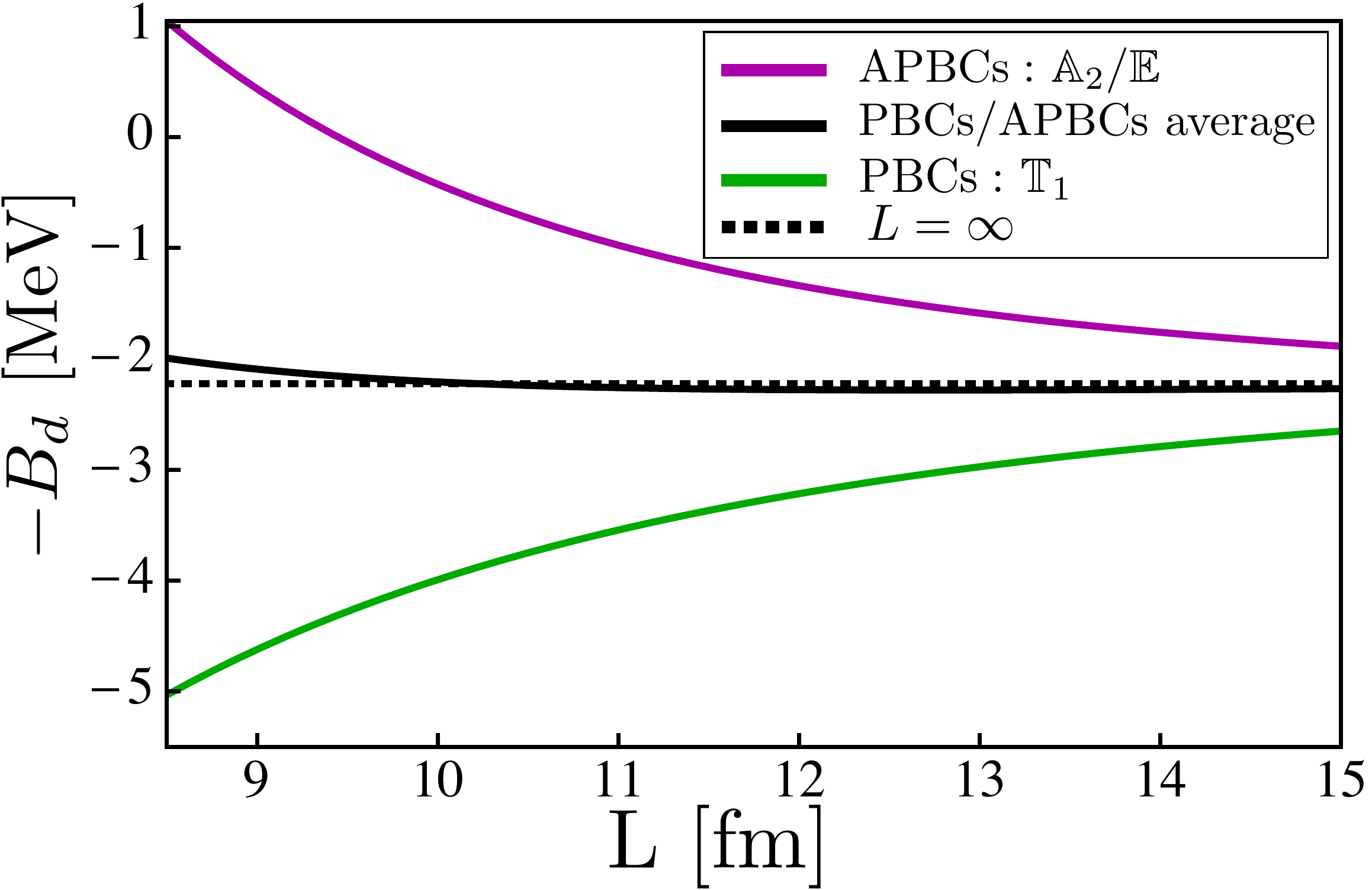}}
\subfigure[]{
\includegraphics[scale=0.3]{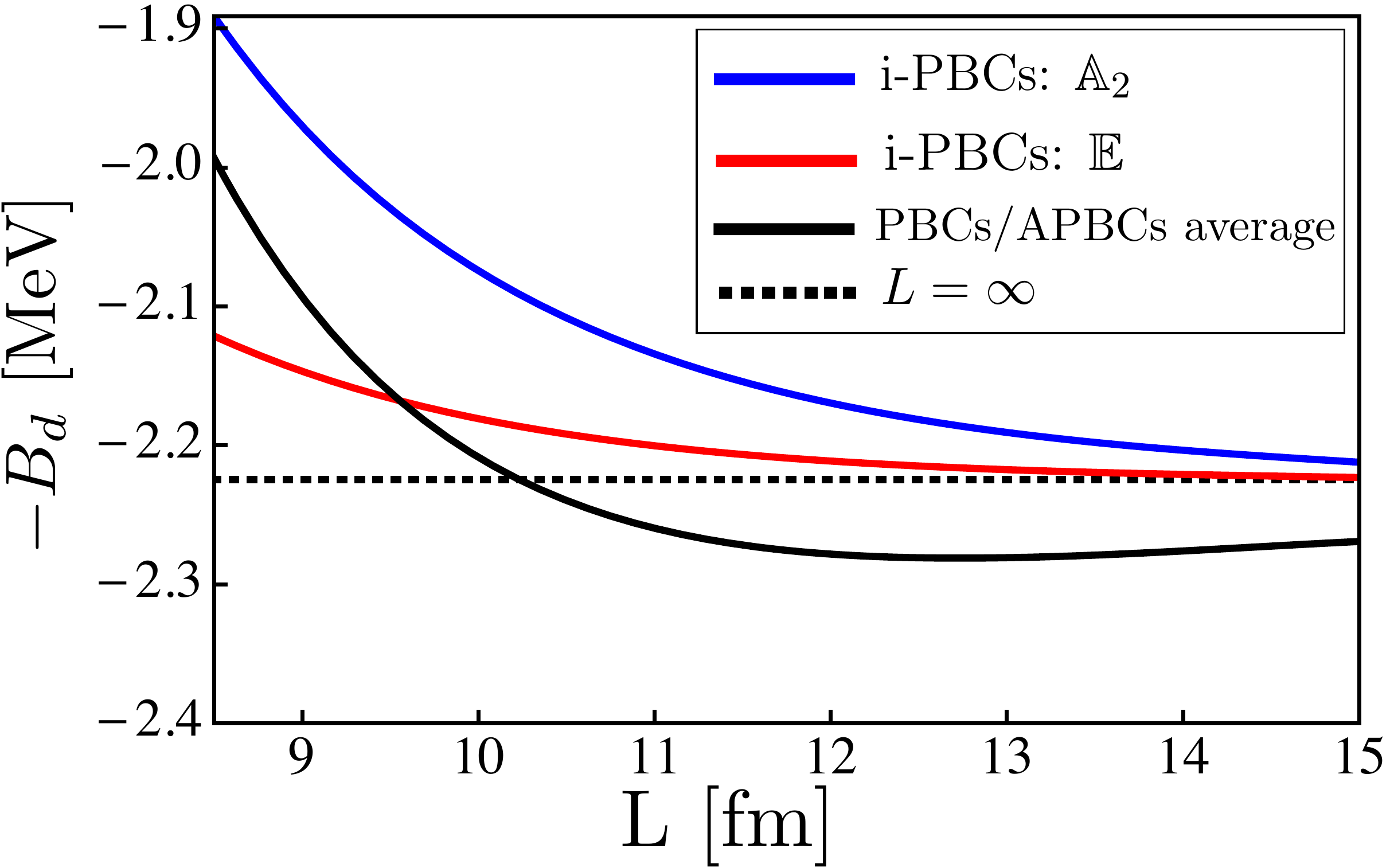}}
\caption{
a) The deuteron binding energy as a function of ${\rm L}$ 
from  PBCs (green curve) and  from APBCs (purple curve). 
The black-solid curve represents the average of these energies. 
b) A closer look at the average in part (a) compared with  energies obtained with i-PBCs,
$\mathbb{A}_2$ (blue curve) and $\mathbb{E}$ (red curve).   The figure is reproduced from Ref. \cite{Briceno:2013hya}.
}
\label{fig:PBC-APBC}
\end{center}
\end{figure}

To appreciate the significance of i-PBCs in reducing the volume corrections to the deuteron binding energy, one can compare the obtained energy levels with that of PBCs and APBCs, as plotted in Fig. \ref{fig:PBC-APBC}(a). As is seen, the FV energies of the deuteron (obtained from the $\mathbb{T}_1$ irrep of the $O_h$ group for the case of PBCs and the $\mathbb{A}_2/\mathbb{E}$ irreps of the $D_{3h}$ group for the case of APBCs) are considerably large such that they make the deuteron in a finite volume either unbound or twice as bound at a volume of $\sim (9~[{\rm fm}])^3$. The energy level obtained by averaging the result of PBCs and APBCs, however, enjoys a large cancellation of the volume effects and gives rise to a determination of the infinite-volume value with an accuracy comparable to that obtained with the i-PBCs, Fig. \ref{fig:PBC-APBC}(b).

The observed improvement of the volume-dependence of the deuteron binding energy can be already understood by studying the S-wave limit of the truncated QC considered in this section,
\begin{eqnarray}
p^*\cot\delta^{{(^3S_1)}}|_{p^*=i\kappa}+\kappa=\sum_{\mathbf{n}\neq\mathbf{0}}
 e^{-i \frac{1}{2} \mathbf{n} \cdot (\bm{\phi}^p-\bm{\phi}^n)}
~\frac{e^{-|\hat{\gamma}\mathbf{n}| \kappa {\rm L}}}{|\hat{\gamma}\mathbf{n}|{\rm L}}
\ \ \ .
\label{S-QC}
\end{eqnarray}
The volume dependence of the deuteron binding energy originates from the right-hand side of this equation. For the twist angles $\bm{\phi}^p=-\bm{\phi}^n \equiv \bm{\phi}=(\frac{\pi}{2},\frac{\pi}{2},\frac{\pi}{2})$,
the first few terms in the summation on the right-hand side of Eq.~(\ref{S-QC}) ($\mathbf{n}^2\leq 3$)
vanish, leaving the leading volume corrections to scale as $\sim e^{-2\kappa{\rm L}}/{\rm L}$.
A lesser 
cancellation occurs in the average of energies obtained with PBCs and APBCs, giving rise to deviations from the infinite-volume energy by terms that scale as $\sim e^{-\sqrt{2}\kappa{\rm L}}/{\rm L}$.

\section{Further Discussions
\label{discussions}}
We conclude by making two comments concerning the generality and the practicality of the use of TBCs in LQCD calculations of two-hadron systems. Firstly, given that the energy eigenvalues in a finite volume are related to scattering amplitudes, in most applications the sizable volume corrections are desirable as they give access to (Minkowski) scattering amplitudes that are otherwise not accessible through (infinite-volume) Euclidean Green's functions \cite{Luscher:1986pf}. Once the scattering amplitudes are constrained, binding energies can be obtained as poles in these amplitudes. Such extraction may however require a parametrization of the energy dependence of the amplitudes (e.g., an effective-range expansion below the t-channel cut). Then the exponential volume-improvement approach already provides an accurate determination of the infinite-volume binding energies in a single volume. Further information regarding scattering parameters can be gained by analyzing the volume effects through the L\"uscher methodology if desired. Performing calculations with different TBCs is advantageous again as it leads to further constraints on the scattering parameters by giving access to extra kinematic points in a finite volume.

It will only be justified, in terms of the associated cost, to perform calculations with multiple TBCs if the generations of new configurations of gauge fields are not required for any set of TBCs. As is pointed out in Ref. \cite{Bedaque:2004ax}, imposing the TBCs only on the valence quarks (partial twisting) is equivalent to the full twisting in the two-nucleon systems up to exponential corrections in volume that have already been neglected in our formalism. This is due to the absence of those s-channel diagrams in nucleon-nucleon interactions that involve intermediate hadrons containing a sea quark. In the scalar sector of QCD the same conclusion has been drawn by the authors of Ref. \cite{Agadjanov:2013kja}. Although in this sector such naively-harmful s-channel diagrams may exist, it can be shown that due to the graded symmetry of the partially-quenched QCD, the same QCs arise as if the full twisting were implemented, up to exponential corrections in volume. These observations ensure that one can enjoy the advantages of the TBCs as discussed in this review and elsewhere in studies of two-hadron interactions with a computational cost that is comparable to, e.g., the boosting method.

\
\

The author would like to thank Ra\'ul Brice\~no, Thomas Luu and Martin Savage for their contribution to the work reviewed in this talk, and for numerous discussions over the topic of this review. This work was supported by the US DOE contract DE-AC05- 06OR23177, under which Jefferson Science Associates, LLC, manages and operates the Jefferson Laboratory, the DFG through SFB/TR 16 and SFG 634, and the DOE grants DE-FG02-00ER41132 and DE-SC0011090.

\end{document}